\documentclass{article}

\usepackage{amssymb}
\usepackage{amsfonts}
\usepackage{amsbsy}
\usepackage[cmtip,all]{xy}
\usepackage{amsmath}
\usepackage{esint}
\usepackage{collectbox}
\usepackage{amscd}
\usepackage{mathrsfs}
\usepackage{amsmath,amsthm}
\usepackage{enumitem}
\usepackage{dsfont}
\usepackage{soul}
\usepackage{mathtools}
\usepackage{comment}
\usepackage{physics}
\usepackage{cite} 
\usepackage[most]{tcolorbox}
\usepackage{stmaryrd}
\usepackage{xcolor}
\definecolor{burgundy}{rgb}{0.5, 0.0, 0.13}
\definecolor{olive}{rgb}{0.50, 0.50, 0.0}
\usepackage[linktocpage=true,colorlinks=true,linkcolor=burgundy,citecolor=black!20!blue,urlcolor=black!30!violet]{hyperref}
\usepackage{accents}
\usepackage{ytableau}
\usepackage{tikz}
\usetikzlibrary{arrows}
\usetikzlibrary{arrows.meta}
\usetikzlibrary{positioning}
\usetikzlibrary{shapes,snakes}
\usetikzlibrary{fit}
\usetikzlibrary{decorations.pathmorphing,decorations.pathreplacing,decorations.markings}
\usetikzlibrary{calc}

\usepackage{pdflscape}
\usepackage{array, longtable}
\newcolumntype{C}{>{$}c<{$}}

\theoremstyle{definition}

\DeclareMathAlphabet{\mathpzc}{OT1}{pzc}{m}{it}

\def\exp{{\rm exp}}

\def\I{{\rm i}}

\def\mod{{\rm mod}}

\def\Tr{{\rm Tr}}

\def\p{\partial}

\def\IC{\mathbb{C}}

\def\IP{\mathbb{P}}

\def\IZ{{\mathbb{Z}}}

\def\fg{\mathfrak{g}}

\def\fl{\mathfrak{l}}

\def\fs{\mathfrak{s}}

\def\fs{\mathfrak{s}}

\def\fQ{\mathfrak{Q}}

\def\lm{\limits}

\hoffset= - 1.0in

\numberwithin{equation}{section}

\DeclareSymbolFont{bbsymbol}{U}{bbold}{m}{n}
\DeclareMathSymbol{\bbzero}{\mathbin}{bbsymbol}{"30}
\DeclareMathSymbol{\bbone}{\mathbin}{bbsymbol}{"31}
\DeclareMathSymbol{\bbtwo}{\mathbin}{bbsymbol}{"32}
\DeclareMathSymbol{\bbthree}{\mathbin}{bbsymbol}{"33}
\DeclareMathSymbol{\bbfour}{\mathbin}{bbsymbol}{"34}
\DeclareMathSymbol{\bbfive}{\mathbin}{bbsymbol}{"35}
\DeclareMathSymbol{\bbsix}{\mathbin}{bbsymbol}{"36}
\DeclareMathSymbol{\bbseven}{\mathbin}{bbsymbol}{"37}
\DeclareMathSymbol{\bbeight}{\mathbin}{bbsymbol}{"38}
\DeclareMathSymbol{\bbnine}{\mathbin}{bbsymbol}{"39}

\tcbset{boxrule=0.6pt,colback=white!97!blue}

\def\myblue{white!40!blue}
\newcommand\sqbox[1]{{
	\setbox0=\hbox{\mbox{$\Box$}}
	\setbox1=\hbox{\mbox{\raisebox{0.35ex}{\tiny #1}}}
	\mbox{\raisebox{-0.2ex}{\rlap{\hbox to \wd0{\hss{\box1}\hss}}\box0}}
}}

\definecolor{red}{rgb}{1,0,0}
\definecolor{orange}{rgb}{1,0.5,0}
\definecolor{violet}{rgb}{0.7,0,1}

\def\be{\begin{eqnarray}}
\def\ee{\end{eqnarray}}

\def\p{\partial}

\def\Tr{{\rm Tr}\,}

\hoffset= - 1.0in         

\begin{document}

\hfill MIPT/TH-16/23

\hfill IITP/TH-15/23

\hfill ITEP/TH-21/23

\vskip 1in
\begin{center}
	
	{\bf\Large{Towards the theory of Yangians}}
	\vskip 0.2in
	\renewcommand{\thefootnote}{\fnsymbol{footnote}}
	{Dmitry Galakhov$^{2,3,4,}$\footnote[2]{e-mail: galakhov@itep.ru},  Alexei Morozov$^{1,2,3,4,}$\footnote[3]{e-mail: morozov@itep.ru} and Nikita Tselousov$^{1,2,4,}$\footnote[4]{e-mail: tselousov.ns@phystech.edu}} 
	\vskip 0.2in 
	\renewcommand{\thefootnote}{\roman{footnote}}
	{\small{ 
			\textit{$^1$MIPT, 141701, Dolgoprudny, Russia}
			\vskip 0 cm 
			\textit{$^2$NRC “Kurchatov Institute”, 123182, Moscow, Russia}
			\vskip 0 cm 
			\textit{$^3$IITP RAS, 127051, Moscow, Russia}
			\vskip 0 cm 
			\textit{$^4$ITEP, Moscow, Russia}
	}}
\end{center}

\vskip 0.2in
\baselineskip 16pt

\centerline{ABSTRACT}

\bigskip

{\footnotesize
	We review the main ideas underlying the emerging theory of Yangians --
	the new type of hidden symmetry in string-inspired models.
	Their classification by quivers is a far-going generalization of
	simple Lie algebras classification by Dynkin diagrams.
	However, this is still a kind of project,
	while a more constructive approach goes through toric Calabi-Yau spaces,
	related supersymmetric systems and the Duistermaat-Heckmann/equivariant integrals
	between the fixed points in the ADHM-like moduli spaces.
	These fixed points are classified by crystals (Young-type diagrams) 
	and Yangian generators describe ``instanton'' transitions between them. Detailed examples will be presented elsewhere.
	
}

\bigskip

\bigskip

\section{Introduction and Discussion}

One of the main goals of string theory \cite{UFN2} is to extend the notion of symmetries,
to make them responsible for most of dynamical properties of physical systems.
This requires a vast extension of the Lie/Noether concept, and such new symmetries
are often referred to as {\it hidden}.
We already know a number of examples -- from {\it integrability} of generic
non-perturbative (functional) integrals \cite{UFN3, GKMMM, Morozov:2022bnz}
and {\it superintegrability} of many of them \cite{Mironov:2022fsr} to more exotic properties like {\it categorical symmetries} \cite{Kong:2022cpy, Bhardwaj:2022yxj, Bah:2022wot} or
{\it differential expansions} \cite{Itoyama:2012fq,Bishler:2020kqw, Morozov:2022aqv} and their {\it defects} \cite{Kononov:2015dda, Morozov:2022zud ,Lanina:2022bxp }
for Wilson loop averages in Chern-Simons theory and knot polynomials.
These hidden symmetries clearly have relation to Lie algebras, but they still remain
somewhat indirect and not fully understood.

\begin{figure}[h!]
	\begin{center}
		\begin{tikzpicture}
			\node[draw = black, rounded corners = 0.1cm](A) at (0,0) {$\begin{array}{c}
					Y_1(\fQ)\\
					\hline
					\mbox{Schurs/Jacks}\\
					\mbox{Times }p_n={\rm ad}_{e_1}^n e_0\\
					\lambda\otimes \lambda'=\bigoplus\lm_{\lambda''}C_{\lambda\lambda'}^{\lambda''}\lambda''\\
					e\sim \Box\otimes\\
					f\sim \langle \Box|\otimes
				\end{array}$};
			\node[draw = black, rounded corners = 0.1cm](B) at (4,0) {$\begin{array}{c}
					Y_2(\fQ)\\
					\hline
					\mbox{DH integrals}\\
					\scalebox{0.6}{$\begin{array}{c}
							E_{\lambda,\lambda+\Box}=\frac{{\rm Eul}_{\lambda}}{{\rm Eul}_{\lambda,\lambda+\Box}}\\
							F_{\lambda+\Box,\lambda}=\frac{{\rm Eul}_{\lambda+\Box}}{{\rm Eul}_{\lambda,\lambda+\Box}}\\
							e(z)|\lambda\rangle=\sum\lm_{\Box\in\lambda^+}\frac{E_{\lambda,\lambda+\Box}}{z-\omega_{\Box}}|\lambda+\Box\rangle\\
							f(z)|\lambda\rangle=\sum\lm_{\Box\in\lambda^-}\frac{F_{\lambda,\lambda-\Box}}{z-\omega_{\Box}}|\lambda-\Box\rangle
						\end{array}$}
				\end{array}$};
			\node[draw = black, rounded corners = 0.1cm](C) at (8,0) {$\begin{array}{c}
					Y_3(\fQ)\\
					\hline
					\mbox{Bootstrap}\\
					\scalebox{0.4}{$e^{(a)}(z)e^{(b)}(w)\cong\varphi_{ab}^Q(z-w)e^{(b)}(w)e^{(a)}(z)$}\\
					\ldots\\
					\scalebox{0.8}{$\varphi_{ab}^{\fQ}(z)=\frac{\prod\lm_{\alpha\in\{a\to b\}}(z+\epsilon_{\alpha})}{\prod\lm_{\beta\in\{b\to a\}}(z-\epsilon_{\beta})}$}
				\end{array}$};
			\node[draw = black, rounded corners = 0.1cm](D) at (12,0) {$\begin{array}{c}
					Y_4(\fQ)\\
					\hline
					\mbox{SQM/QFT}\\
					\mbox{instantons}\\
					\scalebox{0.6}{$\lim\lm_{\tau\to+\infty}\langle\varnothing\otimes\lambda''|e^{-\tau H}|\lambda\otimes\lambda'\rangle$}
				\end{array}$};
			\path (A) edge[<->] (B) (B) edge[<->] (C) (C) edge[<->] (D);
			\node[draw = black, rounded corners = 0.1cm] (E) at (2, 4) {$\begin{array}{c}
					\mbox{ADHM 4d instanton}\\
					\hline
					\sum\lm_{i=1}^{2}\left[B_i,B_i^{\dagger}\right]+II^{\dagger}-J^{\dagger}J=\zeta\bbone\\
					\left[B_1,B_2\right]+IJ=0
				\end{array}$};
			\node[draw = black, rounded corners = 0.1cm] (F) at (10, 4) {$\begin{array}{c}
					\mbox{ADHM 6d ``math'' instantons}\\
					\hline
					\sum\lm_{i=1}^{3}\left[B_i,B_i^{\dagger}\right]+II^{\dagger}-J^{\dagger}J=\zeta\bbone\\
					\left[B_i,B_j\right]=0,\;i,j=1,2,3
				\end{array}$};
			\path (E) edge[->] node[above] {$\scriptstyle +B_3$} (F);
			\draw[->] ([shift={(-0.2,0)}]E.south) to[out=270, in=90] (A.north);
			\draw[->] ([shift={(0.2,0)}]E.south) to[out=270, in=90] ([shift={(-0.2,0)}]B.north);
			\node[draw = black, rounded corners = 0.1cm] (G) at (6,-3) {R-matrix/integrability};
			\draw[->] (A.south) to[out=270, in=180] node[pos=0.3, below left] {\tiny instanton R-matrix?} (G.west);
			\draw[->] (B.south) to[out=270, in=90] ([shift={(-0.5,0)}]G.north);
			\draw[->] (C.south) to[out=270, in=90] node[pos=0.5,below right] {\tiny Coproducts} ([shift={(0.5,0)}]G.north);
			\draw[->] (D.south) to[out=270, in=0] node[pos=0.3, below right] {\tiny gauge/Bethe} (G.east);
			\draw[->, decorate, decoration={snake, segment length=2mm, amplitude=0.5mm}] (F.south) to[out=270,in=90] node(FtoB) [pos=0.1, below right] {\tiny Strictly speaking} ([shift={(0.2,0)}]B.north);
			\node[below] at (FtoB) {\tiny variety is singular};
			\node[draw = black, rounded corners = 0.1cm] (H) at (2,11) {toric CY${}_3$};
			\node[draw = black, rounded corners = 0.1cm] (I) at (6,10) {$\begin{array}{c}
					\mbox{Periodic quiver $\hat \fQ$}\\
					\mbox{on a torus (lattice)}\\
					\hline\\
					\begin{tikzpicture}[scale=0.5]
						\tikzset{arr1/.style={
								postaction={decorate},
								decoration={markings, mark= at position 0.6 with {\arrow{stealth}}},
								thick,black!30!red}}
						\tikzset{arr2/.style={
								postaction={decorate},
								decoration={markings, mark= at position 0.6 with {\arrow{stealth}}},
								thick,\myblue}}
						\tikzset{arr3/.style={
								postaction={decorate},
								decoration={markings, mark= at position 0.6 with {\arrow{stealth}}},
								thick,black!60!green}}
						\foreach \x in {0,1,...,3}
						\foreach \y in {0,1,...,2}
						{
							\draw[arr1] (\x + 0.5*\y, 0.866025*\y) -- (\x + 0.5*\y + 0.5, 0.866025*\y + 0.866025);
							\draw[arr2] (\x + 1 + 0.5*\y, 0.866025*\y) -- (\x + 0.5*\y, 0.866025*\y);
							\draw[arr3] (\x + 0.5*\y + 0.5, 0.866025*\y + 0.866025) -- (\x + 1 + 0.5*\y, 0.866025*\y);
						}
						\foreach \x in {0,1,...,3}
						\foreach \y in {3}
						{
							\draw[arr2] (\x + 1 + 0.5*\y, 0.866025*\y) -- (\x + 0.5*\y, 0.866025*\y);
						}
						\foreach \x in {4}
						\foreach \y in {0,1,...,2}
						{
							\draw[arr1] (\x + 0.5*\y, 0.866025*\y) -- (\x + 0.5*\y + 0.5, 0.866025*\y + 0.866025);
						}
						\foreach \x in {0,1,...,4}
						\foreach \y in {0,1,...,3}
						{
							\draw[fill=white] (\x + 0.5*\y, 0.866025*\y) circle (0.1);
						}
					\end{tikzpicture}
				\end{array}$};
			\node[draw = black, rounded corners = 0.1cm] (J) at (12,10) {$\begin{array}{c}
					\mbox{Quiver $\fQ$ }\begin{array}{c}
						\begin{tikzpicture}[scale=0.6]
							\draw[rounded corners = 0cm, thick, black!30!red, postaction={decorate},decoration={markings,
								mark= at position 0.8 with {\arrow{stealth}}}] (0,0) to[out=60,in=0] (0,1) to[out=180,in=120] (0,0);
							\begin{scope}[rotate=120]
								\draw[rounded corners = 0cm, thick, \myblue, postaction={decorate},decoration={markings,
									mark= at position 0.8 with {\arrow{stealth}}}] (0,0) to[out=60,in=0] (0,1) to[out=180,in=120] (0,0);
							\end{scope}
							\begin{scope}[rotate=240]
								\draw[rounded corners = 0cm, thick, black!60!green, postaction={decorate},decoration={markings,
									mark= at position 0.8 with {\arrow{stealth}}}] (0,0) to[out=60,in=0] (0,1) to[out=180,in=120] (0,0);
							\end{scope}
							\draw[fill=white] (0,0) circle (0.1);
						\end{tikzpicture}
					\end{array}\\
					\mbox{+superpotential}\\
					W=\sum\lm_{f\in{\rm faces}(\hat \fQ)}(-1)^{\rm ori}\Tr\prod\lm_{\ell\in\p f}^{\curvearrowleft}B_{\ell}
				\end{array}$};
			\node[draw = black, rounded corners = 0.1cm] (K) at (10,6.8) {$\begin{array}{r}
					\mbox{D-term: }\sum\lm_{\alpha\in\{*\to a\}}\phi_{\alpha}\phi_{\alpha}^{\dagger}-\sum\lm_{\beta\in\{a\to *\}}\phi_{\beta}^{\dagger}\phi_{\beta}=\zeta_a\,\bbone,\;a\in \fQ_0\\
					\mbox{F-term: }\p_{\phi_{\alpha}}W=0,\;\alpha\in \fQ_1
				\end{array}$};
			\draw[->] (I.east) -- (J.west);
			\draw[->] (J.south) to[out=270, in=90] (K.north);
			\draw[->] (K.south) to[out=270, in=90] node[right] {$\scriptstyle W_0=\Tr B_3\left[B_1,B_2\right]$} (F.north);
			\node[draw = black, rounded corners = 0.1cm] (L) at (2,8.5) {$\begin{array}{c}
					\mbox{Framings}\\
					\hline
					\begin{tikzpicture}
						\draw[postaction={decorate},
						decoration={markings, mark= at position 0.65 with {\arrow{stealth}}}] (0,0) to[out=10,in=170] (1.5,0);
						\draw[postaction={decorate},
						decoration={markings, mark= at position 0.65 with {\arrow{stealth}}}] (1.5,0) to[out=190,in=-10] (0,0);
						\draw[fill=\myblue,rounded corners = 0cm] (-0.1,-0.1) -- (-0.1,0.1) -- (0.1,0.1) -- (0.1,-0.1) -- cycle;
						\draw[dashed, fill=white] (1.5,0) circle (0.25);
						\node at (1.5,0) {$\fQ$};
					\end{tikzpicture}\\
					\mbox{D-brane wraps}\\
					\mbox{some divisor}\\
					\mbox{of toric CY${}_3$}\\
					\mbox{(Representations)}
				\end{array}$};
			\draw[->] (H.east) to[out=0, in=180] (I.west);
			\draw[->] (H.south) to[out=270,in=90] (L.north);
			\draw (L.south) to[out=270, in=90] (3,6);
			\draw[->] (3,6) to[out=270, in=90] (E.north);
			\draw (K.west) to[out=180,in=0] (3,6);
			\draw[fill=white] (3,6) circle (0.05);
			\node[left] at (2,6) {$\begin{array}{c}
					\scriptstyle W=W_0+\Tr B_3IJ\\
					\scriptstyle B_3=0\; \left(\begin{array}{c}
						\mbox{\tiny ``Lagrange''}\\
						\mbox{\tiny multiplier}
					\end{array}\right)
				\end{array}$};
			\node[draw = black, rounded corners = 0.1cm](M) at (10,-4.3) {$\begin{array}{c}
					\mbox{Cohomological}\\
					\mbox{Hall algebra}\\
					\mbox{of Kontsevich-Soibelman}
				\end{array}$};
			\draw[->] (M.north) to[out=90,in=270] ([shift={(1,0)}]C.south);
			\draw[->, dashed] ([shift={(1,0)}]M.north) to[out=90,in=270] ([shift={(1,0)}]D.south);
			\node[draw = black, rounded corners = 0.1cm] (N) at (12,-7) {$\begin{array}{c}
					\mbox{\bf ???}\\
					\mbox{Moduli space of}\\
					\mbox{$\zeta_a$-parameters}\\
					\hline
					\mbox{Stability chambers}\\
					\hline
					\mbox{Wall-crossing}
				\end{array}$};
			\draw[->] (12.5,6.9) to[out=270,in=90] (13.7,0) -- (13.7,-4.3) to[out=270,in=90] ([shift={(1,0)}]N.north);
			\node at (10.1,-7) {$\in$};
			\node[draw = black, rounded corners = 0.1cm] (P) at (7.5,-7.5) {$\begin{array}{c}
					\mbox{Mutations}\\
					\hline
					\mbox{(Seiberg duality)}\\
					\mathscr{M}(\fQ,\zeta,d)\cong\mathscr{M}(\fQ',\zeta',d')\\
					\scalebox{0.5}{$\begin{array}{c}
							\begin{tikzpicture}
								\begin{scope}
									\draw[postaction={decorate},
									decoration={markings, mark= at position 0.65 with {\arrow{stealth}}}] (1,0) to[out=105,in=315] (0.5, 0.866025);
									\draw[postaction={decorate},
									decoration={markings, mark= at position 0.65 with {\arrow{stealth}}}] (0.5, 0.866025) to[out=285,in=135] (1,0);
									\draw[postaction={decorate},
									decoration={markings, mark= at position 0.85 with {\arrow{stealth}}},rounded corners = 0cm] (1,0) to[out=330,in=270] (1.3,0) to[out=90,in=30] (1,0);
								\end{scope}
								\begin{scope}[rotate=60]
									\draw[postaction={decorate},
									decoration={markings, mark= at position 0.65 with {\arrow{stealth}}}] (1,0) to[out=105,in=315] (0.5, 0.866025);
									\draw[postaction={decorate},
									decoration={markings, mark= at position 0.65 with {\arrow{stealth}}}] (0.5, 0.866025) to[out=285,in=135] (1,0);
									\draw[postaction={decorate},
									decoration={markings, mark= at position 0.85 with {\arrow{stealth}}},rounded corners = 0cm] (1,0) to[out=330,in=270] (1.3,0) to[out=90,in=30] (1,0);
								\end{scope}
								\begin{scope}[rotate=120]
									\draw[postaction={decorate},
									decoration={markings, mark= at position 0.65 with {\arrow{stealth}}}] (1,0) to[out=105,in=315] (0.5, 0.866025);
									\draw[postaction={decorate},
									decoration={markings, mark= at position 0.65 with {\arrow{stealth}}}] (0.5, 0.866025) to[out=285,in=135] (1,0);
									\draw[postaction={decorate},
									decoration={markings, mark= at position 0.85 with {\arrow{stealth}}},rounded corners = 0cm] (1,0) to[out=330,in=270] (1.3,0) to[out=90,in=30] (1,0);
								\end{scope}
								\begin{scope}[rotate=180]
									\draw[postaction={decorate},
									decoration={markings, mark= at position 0.65 with {\arrow{stealth}}}] (1,0) to[out=105,in=315] (0.5, 0.866025);
									\draw[postaction={decorate},
									decoration={markings, mark= at position 0.65 with {\arrow{stealth}}}] (0.5, 0.866025) to[out=285,in=135] (1,0);
								\end{scope}
								\begin{scope}[rotate=240]
									\draw[postaction={decorate},
									decoration={markings, mark= at position 0.65 with {\arrow{stealth}}}] (1,0) to[out=105,in=315] (0.5, 0.866025);
									\draw[postaction={decorate},
									decoration={markings, mark= at position 0.65 with {\arrow{stealth}}}] (0.5, 0.866025) to[out=285,in=135] (1,0);
									\draw[postaction={decorate},
									decoration={markings, mark= at position 0.85 with {\arrow{stealth}}},rounded corners = 0cm] (1,0) to[out=330,in=270] (1.3,0) to[out=90,in=30] (1,0);
								\end{scope}
								\begin{scope}[rotate=300]
									\draw[postaction={decorate},
									decoration={markings, mark= at position 0.65 with {\arrow{stealth}}}] (1,0) to[out=105,in=315] (0.5, 0.866025);
									\draw[postaction={decorate},
									decoration={markings, mark= at position 0.65 with {\arrow{stealth}}}] (0.5, 0.866025) to[out=285,in=135] (1,0);
								\end{scope}
								\foreach \i in {0, 60, 120, 180, 240, 300}
								{
									\begin{scope}[rotate=\i]
										\draw[fill=white] (1,0) circle (0.08);
									\end{scope}
								}
							\end{tikzpicture}
						\end{array}$}\leftrightarrow\scalebox{0.5}{$\begin{array}{c}
							\begin{tikzpicture}
								\begin{scope}
									\draw[postaction={decorate},
									decoration={markings, mark= at position 0.65 with {\arrow{stealth}}}] (1,0) to[out=105,in=315] (0.5, 0.866025);
									\draw[postaction={decorate},
									decoration={markings, mark= at position 0.65 with {\arrow{stealth}}}] (0.5, 0.866025) to[out=285,in=135] (1,0);
								\end{scope}
								\begin{scope}[rotate=60]
									\draw[postaction={decorate},
									decoration={markings, mark= at position 0.65 with {\arrow{stealth}}}] (1,0) to[out=105,in=315] (0.5, 0.866025);
									\draw[postaction={decorate},
									decoration={markings, mark= at position 0.65 with {\arrow{stealth}}}] (0.5, 0.866025) to[out=285,in=135] (1,0);
									\draw[postaction={decorate},
									decoration={markings, mark= at position 0.85 with {\arrow{stealth}}},rounded corners = 0cm] (1,0) to[out=330,in=270] (1.3,0) to[out=90,in=30] (1,0);
								\end{scope}
								\begin{scope}[rotate=120]
									\draw[postaction={decorate},
									decoration={markings, mark= at position 0.65 with {\arrow{stealth}}}] (1,0) to[out=105,in=315] (0.5, 0.866025);
									\draw[postaction={decorate},
									decoration={markings, mark= at position 0.65 with {\arrow{stealth}}}] (0.5, 0.866025) to[out=285,in=135] (1,0);
									\draw[postaction={decorate},
									decoration={markings, mark= at position 0.85 with {\arrow{stealth}}},rounded corners = 0cm] (1,0) to[out=330,in=270] (1.3,0) to[out=90,in=30] (1,0);
								\end{scope}
								\begin{scope}[rotate=180]
									\draw[postaction={decorate},
									decoration={markings, mark= at position 0.65 with {\arrow{stealth}}}] (1,0) to[out=105,in=315] (0.5, 0.866025);
									\draw[postaction={decorate},
									decoration={markings, mark= at position 0.65 with {\arrow{stealth}}}] (0.5, 0.866025) to[out=285,in=135] (1,0);
								\end{scope}
								\begin{scope}[rotate=240]
									\draw[postaction={decorate},
									decoration={markings, mark= at position 0.65 with {\arrow{stealth}}}] (1,0) to[out=105,in=315] (0.5, 0.866025);
									\draw[postaction={decorate},
									decoration={markings, mark= at position 0.65 with {\arrow{stealth}}}] (0.5, 0.866025) to[out=285,in=135] (1,0);
								\end{scope}
								\begin{scope}[rotate=300]
									\draw[postaction={decorate},
									decoration={markings, mark= at position 0.65 with {\arrow{stealth}}}] (1,0) to[out=105,in=315] (0.5, 0.866025);
									\draw[postaction={decorate},
									decoration={markings, mark= at position 0.65 with {\arrow{stealth}}}] (0.5, 0.866025) to[out=285,in=135] (1,0);
								\end{scope}
								\foreach \i in {0, 60, 120, 180, 240, 300}
								{
									\begin{scope}[rotate=\i]
										\draw[fill=white] (1,0) circle (0.08);
									\end{scope}
								}
							\end{tikzpicture}
						\end{array}$}
				\end{array}$};
			\node[draw = black, rounded corners = 0.1cm](O) at (1.5,-6){$\begin{array}{c}
					\mbox{Crystals/tessalations/}\\
					\mbox{level lines on Voronoi lattice}\\
					\hline
					\begin{array}{c}
						\begin{tikzpicture}[scale=0.4]
							\draw (0.110616,2.59346) -- (0.698401,2.34346) (0.110616,2.59346) -- (-0.698401,2.41182) (0.698401,1.3924) -- (0.698401,2.34346) (0.698401,1.3924) -- (1.28619,1.1424) (0.698401,1.3924) -- (-0.110616,1.21077) (0.698401,2.34346) -- (-0.110616,2.16182) (1.28619,0.191346) -- (1.28619,1.1424) (1.28619,0.191346) -- (1.87397,-0.0586539) (1.28619,0.191346) -- (0.477169,0.00971044) (1.28619,1.1424) -- (0.477169,0.960767) (1.87397,-0.0586539) -- (2.46176,-0.308654) (1.87397,-0.0586539) -- (1.06495,-0.24029) (2.46176,-1.25971) -- (2.46176,-0.308654) (2.46176,-1.25971) -- (1.65274,-1.44135) (2.46176,-0.308654) -- (1.65274,-0.49029) (-0.698401,1.46077) -- (-0.698401,2.41182) (-0.698401,1.46077) -- (-0.110616,1.21077) (-0.698401,1.46077) -- (-1.50742,1.27913) (-0.698401,2.41182) -- (-0.110616,2.16182) (-0.110616,1.21077) -- (-0.110616,2.16182) (-0.110616,1.21077) -- (0.477169,0.960767) (-0.110616,1.21077) -- (-0.919633,1.02913) (0.477169,0.00971044) -- (0.477169,0.960767) (0.477169,0.00971044) -- (1.06495,-0.24029) (0.477169,0.00971044) -- (-0.331848,-0.171925) (0.477169,0.960767) -- (-0.331848,0.779131) (1.06495,-1.19135) -- (1.06495,-0.24029) (1.06495,-1.19135) -- (1.65274,-1.44135) (1.06495,-1.19135) -- (0.255938,-1.37298) (1.06495,-0.24029) -- (1.65274,-0.49029) (1.06495,-0.24029) -- (0.255938,-0.421925) (1.65274,-1.44135) -- (1.65274,-0.49029) (-1.50742,0.328075) -- (-1.50742,1.27913) (-1.50742,0.328075) -- (-0.919633,0.0780748) (-1.50742,0.328075) -- (-2.31644,0.146439) (-1.50742,1.27913) -- (-0.919633,1.02913) (-0.919633,-0.872982) -- (-0.919633,0.0780748) (-0.919633,-0.872982) -- (-0.331848,-1.12298) (-0.919633,-0.872982) -- (-1.72865,-1.05462) (-0.919633,0.0780748) -- (-0.919633,1.02913) (-0.919633,0.0780748) -- (-0.331848,-0.171925) (-0.919633,0.0780748) -- (-1.72865,-0.103561) (-0.919633,1.02913) -- (-0.331848,0.779131) (-0.331848,-1.12298) -- (-0.331848,-0.171925) (-0.331848,-1.12298) -- (0.255938,-1.37298) (-0.331848,-0.171925) -- (-0.331848,0.779131) (-0.331848,-0.171925) -- (0.255938,-0.421925) (0.255938,-1.37298) -- (0.255938,-0.421925) (-2.31644,-0.804617) -- (-2.31644,0.146439) (-2.31644,-0.804617) -- (-1.72865,-1.05462) (-2.31644,0.146439) -- (-1.72865,-0.103561) (-1.72865,-1.05462) -- (-1.72865,-0.103561);
						\end{tikzpicture}
					\end{array}\leftrightarrow
					\begin{array}{c}
						\begin{tikzpicture}[scale=0.4]
							\draw[thick, gray] (-2.3094,0.)--(-2.02073,-0.5)(-2.3094,0.)--(-2.02073,0.5)(-2.3094,1.)--(-2.02073,0.5)(-2.3094,1.)--(-2.02073,1.5)(-2.3094,2.)--(-2.02073,1.5)(-2.3094,2.)--(-2.02073,2.5)(-2.02073,-0.5)--(-1.44338,-0.5)(-2.02073,2.5)--(-1.44338,2.5)(-1.44338,-1.5)--(-1.1547,-2.)(-1.44338,-1.5)--(-1.1547,-1.)(-1.44338,-0.5)--(-1.1547,0.)(-1.44338,0.5)--(-1.1547,0.)(-1.44338,0.5)--(-1.1547,1.)(-1.44338,2.5)--(-1.1547,2.)(-1.1547,-2.)--(-0.57735,-2.)(-1.1547,-1.)--(-0.57735,-1.)(-1.1547,1.)--(-0.57735,1.)(-1.1547,2.)--(-0.57735,2.)(-0.57735,-2.)--(-0.288675,-1.5)(-0.57735,-1.)--(-0.288675,-1.5)(-0.57735,1.)--(-0.288675,1.5)(-0.57735,2.)--(-0.288675,1.5)(0.288675,1.5)--(0.57735,1.)(0.288675,1.5)--(0.57735,2.)(0.57735,1.)--(1.1547,1.)(0.57735,2.)--(1.1547,2.)(1.1547,0.)--(1.44338,-0.5)(1.1547,0.)--(1.44338,0.5)(1.1547,1.)--(1.44338,0.5)(1.1547,2.)--(1.44338,2.5)(1.44338,-0.5)--(2.02073,-0.5)(1.44338,2.5)--(2.02073,2.5)(2.02073,-0.5)--(2.3094,-1.)(2.02073,2.5)--(2.3094,2.)(2.3094,-1.)--(2.88675,-1.)(2.3094,2.)--(2.88675,2.)(2.88675,-1.)--(3.17543,-0.5)(2.88675,0.)--(3.17543,-0.5)(2.88675,0.)--(3.17543,0.5)(2.88675,1.)--(3.17543,0.5)(2.88675,1.)--(3.17543,1.5)(2.88675,2.)--(3.17543,1.5);
							\draw[thick, \myblue] (-2.3094,-1.)--(-2.02073,-1.5)(-2.3094,-1.)--(-2.02073,-0.5)(-2.02073,-1.5)--(-1.44338,-1.5)(-2.02073,-0.5)--(-1.44338,-0.5)(-1.44338,-1.5)--(-1.1547,-1.)(-1.44338,-0.5)--(-1.1547,0.)(-1.44338,0.5)--(-1.1547,0.)(-1.44338,0.5)--(-1.1547,1.)(-1.1547,-1.)--(-0.57735,-1.)(-1.1547,1.)--(-0.57735,1.)(-0.57735,-1.)--(-0.288675,-1.5)(-0.57735,0.)--(-0.288675,-0.5)(-0.57735,0.)--(-0.288675,0.5)(-0.57735,1.)--(-0.288675,1.5)(-0.57735,2.)--(-0.288675,1.5)(-0.57735,2.)--(-0.288675,2.5)(-0.288675,-1.5)--(0.288675,-1.5)(-0.288675,-0.5)--(0.288675,-0.5)(-0.288675,0.5)--(0.288675,0.5)(-0.288675,2.5)--(0.288675,2.5)(0.288675,-1.5)--(0.57735,-2.)(0.288675,-0.5)--(0.57735,0.)(0.288675,0.5)--(0.57735,0.)(0.288675,1.5)--(0.57735,1.)(0.288675,1.5)--(0.57735,2.)(0.288675,2.5)--(0.57735,2.)(0.57735,-2.)--(1.1547,-2.)(0.57735,1.)--(1.1547,1.)(1.1547,-2.)--(1.44338,-1.5)(1.1547,0.)--(1.44338,-0.5)(1.1547,0.)--(1.44338,0.5)(1.1547,1.)--(1.44338,0.5)(1.44338,-1.5)--(2.02073,-1.5)(1.44338,-0.5)--(2.02073,-0.5)(2.02073,-1.5)--(2.3094,-2.)(2.02073,-0.5)--(2.3094,-1.)(2.3094,-2.)--(2.88675,-2.)(2.3094,-1.)--(2.88675,-1.)(2.88675,-2.)--(3.17543,-1.5)(2.88675,-1.)--(3.17543,-1.5);
							\draw[thick,orange] (-0.57735,0.)--(-0.288675,-0.5)(-0.57735,0.)--(-0.288675,0.5)(-0.288675,-0.5)--(0.288675,-0.5)(-0.288675,0.5)--(0.288675,0.5)(0.288675,-0.5)--(0.57735,0.)(0.288675,0.5)--(0.57735,0.);
							\foreach \x/\y in {1.73205/2., 2.59808/1.5, 0.866025/1.5, 1.73205/1., 2.59808/0.5, -1.73205/2., -0.866025/1.5, 1.73205/0., 2.59808/-0.5, -1.73205/1., -1.73205/0., -0.866025/-1.5}
							{
								\node[gray] at (\x,\y) {\tiny \bf 0};
							}
							\foreach \x/\y in {0./2., 0./1., 0.866025/0.5, -0.866025/0.5, 0.866025/-0.5, 1.73205/-1., 2.59808/-1.5, -0.866025/-0.5, 0./-1., 0.866025/-1.5, -1.73205/-1.}
							{
								\node[\myblue] at (\x,\y) {\tiny \bf 1};
							}
							\foreach \x/\y in {0./0.}
							{
								\node[orange] at (\x,\y) {\tiny \bf 2};
							}
						\end{tikzpicture}
					\end{array}
				\end{array}$};
			\draw[->] ([shift={(0,0.5)}]I.west) to[out=180,in=0] (2,12) to[out=180,in=90] (-2,4) -- (-2,-3) to[out=270,in=90] ([shift={(-1,0)}]O.north);
			\draw[->] (O.north) to[out=90,in=270] node[pos=0.8,right] {\tiny Diagrams} ([shift={(1,0)}]A.south);
			\draw[->] (O.north) to[out=90,in=270] ([shift={(-1,0)}]B.south);
			\draw[->] ([shift={(2,0)}]O.north) -- ([shift={(2,2)}]O.north) to[out=90,in=270] ([shift={(-1.5,0)}]C.south);
			\draw[->, thick, dashed] ([shift={(0,1)}]N.west) to[out=180,in=0] ([shift={(0,1)}]O.east);
		\end{tikzpicture}
		\caption{Approaches to the quiver Yangians}\label{fig}
	\end{center}
\end{figure}
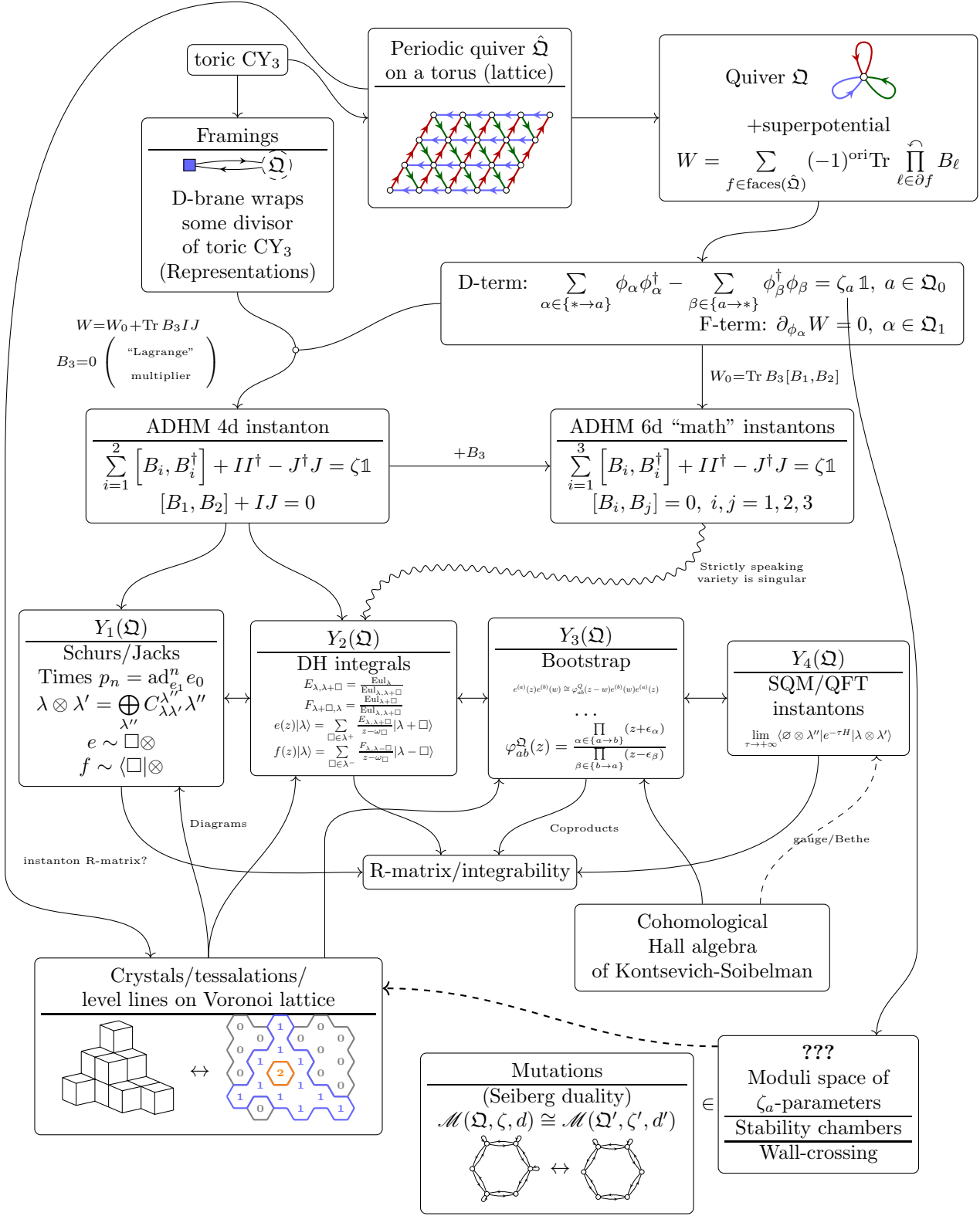

Much closer to direct deformation of Lie structure are  {\it Yangians} \cite{Harvey:1996gc, Tsymbaliuk:2014fvq,Prochazka:2015deb,Yamazaki:2022cdg, Li:2020rij, Galakhov:2020vyb} --
and naturally they get more and more attention in recent years \cite{Bao:2022jhy,2019arXiv191106666U,Kolyaskin:2022tqi,wang2020affine, Gaberdiel:2017dbk}.
Though original motivation, which we begin from in this introduction,
is very much in the spirit of other {\it hidden} symmetries,
it is clear that Cartan-like formal description,
similar to that of simple Lie algebras, is also available \cite{Galakhov:2021xum, Yamazaki:2022cdg, Li:2020rij, Li:2023zub} (see s.\ref{formal definition}), where conventional Dynkin diagrams are generalized to quivers (see s.\ref{def quiver}).
However, even in this case there is still a big way towards understanding of
representation theory of Yangians, and more constructive ``physical'' approaches
continue to dominate \cite{Li:2020rij,Galakhov:2020vyb} -- as we discuss in the sections \ref{Yangain from quiver reps} and \ref{Yangain from QFT} and summarize
in the Figure \ref{fig}.

In this short paper we outlined the main ideas used in description of Yangians and their representations. These ideas are four different points of view on the same quiver Yangian, that corresponds to boxes $Y_1(\fQ),Y_2(\fQ),Y_3(\fQ),Y_4(\fQ)$ on the Figure \ref{fig} and sections \ref{Jacks}, \ref{Yangain from quiver reps}, \ref{Yangain from quiver}, \ref{Yangain from QFT} respectively.
Detailed examples and further review of existing literature will be presented elsewhere. Further deformations to DIM algebras \cite{DIM1, DIM2, Noshita:2021dgj} and associated brane/network  models \cite{Mironov:2016yue}
can use the same patterns, along with the matrix model \cite{Mironov:2010zs} and CFT (free field)
\cite{GMMOS,Morozov:2022ocp, Kolyaskin:2022tqi, Chistyakova:2021yyd} techniques.

\bigskip

\section{Yangian from Schur-Jack family of polynomials}
\label{Jacks}
In this section we discuss the first appearance of quiver Yangian algebras through the family of Schur-Jack polynomials that correspond to rectangular $Y_1(\fQ)$ of the Figure \ref{fig}.

Schur polynomials $S_{R}$ form a distinguished basis in the space of homogeneous polynomials of variables $p_a$, $a = 1,2,3, \ldots, \infty$ . One of the crucial properties of the Schur polynomials is that they form a character ring of $GL_{N}$:
\begin{equation}
	R \otimes R' = \sum_{R''} N^{R''}_{RR'} R'' \hspace{15mm} \Rightarrow \hspace{15mm} S_R \cdot S_{R'} = \sum_{R''} N^{R''}_{RR'} S_{R''}
\end{equation}
where $N^{R''}_{RR'}$ are the famous Littlewood-Richardson coefficients and $R, R^{'}, R^{''}$ are the Young diagrams:
\begin{equation}
	\ytableausetup{boxsize = 0.8em}
	R = \begin{ytableau}
		\ & \ & \ & \ & \ \\
		\ & \ & \ \\
		\ & \ \\ 
	\end{ytableau} \hspace{15mm} R = [R_1, R_2, \ldots, R_{l(R)}], \ \text{where} \ R_i \in \mathbb{Z} \ \text{and} \ \ R_1 \geqslant R_2 \geqslant \ldots \geqslant R_{l(R)} > 0 
\end{equation}
The numbers $p(n)$ of Young diagrams with $n$ boxes are nicely collected in the following generating function:
\begin{equation}
	\prod_{k = 1} \frac{1}{1 - q^k} = \sum_{n = 0} p(n) \, q^n = 1 + q + 2 q^2 + 3 q^3 + 5 q^4 + 7 q^5 + \ldots
\end{equation}
Orthogonality property for Schur polynomials can be represented as the Cauchy identity:
\begin{equation}
	\sum_{R} S_{R} \{p\} S_{R^{'}} \{ \bar{p} \} = \exp \left\{ \sum_{k = 1}^{\infty} \frac{p_k \, \bar{p}_k}{k} \right\}
\end{equation}

An avatar of Schur-Weyl duality between linear and symmetric groups \cite{} can be formulated in terms of commuting set of cut-and-join operators \cite{MMN}:
\begin{equation}
	\hat{W}_{\Delta} \, S_{R} = \phi_{R}(\Delta) \, S_{R}
\end{equation}
where $\phi_{R}(\Delta)$ are properly normalized characters of symmetric groups. The simplest non-trivial cut-and-join operator $\hat{W}_{[2]}$ actually defines the Schur functions as the set of own eigenfunctions. In other words, all the Schur polynomials encoded in the following nicely looking operator:
\begin{equation}
	\hat{W}_{[2]} = \frac{1}{2} \sum_{a, b = 1}^{\infty} \left[ a b \, p_{a + b} \frac{\partial^2}{\partial p_a \partial p_b} + (a + b) \, p_a \, p_b \frac{\partial}{\partial p_{a+b}} \right]
\end{equation}
An important fact for our presentation is that the commutative algebra of cut-and-join operators can be extended to the full non-commutative $W_{1 + \infty}$ algebra \cite{AFMO}\footnote{Another avatar for these algebras are so called vertex operator algebras (VOA). On recent developments see \cite{Prochazka:2017qum, Gaiotto:2020dsq, Cheng:2022rqr, Gaiotto:2023ynn, Kimura:2023bxy} and references therein.}, that is a special case of affine Yangian $Y(\hat{\mathfrak{gl}}_1)$ \cite{Tsymbaliuk:2014fvq, Prochazka:2015deb}. Remarkably this Yangian is generated by multiple commutators of the small set of simple operators :
\begin{itemize}
	\item operator of multiplication on the first $p$-variable $p_1$:
	\begin{equation}
		p_1 \cdot S_{R} = \sum_{\Box \in R^{+}} S_{R + \Box}
	\end{equation}
	\item derivative with respect to $p_1$:
	\begin{equation}
		\frac{\partial}{\partial p_1}  S_{R} = \sum_{\Box \in R^{-}} S_{R - \Box} 
	\end{equation}
	\item diagonal operator $\hat{W}_{[2]}$:
	\begin{equation}
		\hat{W}_{[2]}\, S_{R} = \left( \sum_{\Box \in R} j_{\Box} - i_{\Box} \right) \cdot S_{R}
	\end{equation}
\end{itemize}
where $i_{\Box}, j_{\Box}$ are the two coordinates of the box in the Young diagram. The notation $R^{\pm}$ means the positions of the boxes outside/inside the Young diagram where one can add/remove a box in a way that $R \pm \Box$ is still a Young diagram. 
\bigskip

Miraculously almost the whole picture lifts to the level of $\beta$-deformation \cite{Morozov:2012dz}. The Schurs $S_{R}$ become Jack polynomials $J_{R}$ \cite{Macdonald}:
\begin{equation}
	J_R \cdot J_{R'} = \sum_{R''} {\cal N}^{R''}_{RR'}(\beta)  \, J_{R''}
\end{equation}
Orthogonality property survives $\beta$-deformation:
\begin{equation}
	\sum_{R} \frac{J_{R} \{p\} J_{R^{'}} \{ \bar{p} \}}{||J_{R}||^2}  = \exp \left\{ \sum_{k = 1}^{\infty} \frac{ \beta \cdot p_k \, \bar{p}_k}{k} \right\}
\end{equation}
and cut-and-join operator undergoes simple deformation:
\begin{equation}
	\hat{W}_{[2]}^{\beta} = \frac{1}{2} \sum_{a, b = 1}^{\infty} \left[ a b \, p_{a + b} \frac{\partial^2}{\partial p_a \partial p_b} + \beta (a + b) \, p_a \, p_b \frac{\partial}{\partial p_{a+b}}\right] + \frac{(1-\beta)}{2} \sum_{a = 1}^{\infty} (a-1)a \, p_a \frac{\partial}{\partial p_a}
\end{equation}
In $\beta$-deformed case this operator corresponds to the integrable Hamiltonian of quantum many-body system \cite{Awata2000, Mironov:2023zwi}. Triple of operators generate the algebra affine Yangian $Y(\hat{\mathfrak{gl}}_1)$:
\begin{equation}
	e_0 = p_1, \hspace{10mm} \psi_3 = \hat{W}_{[2]}^{\beta}, \hspace{10mm} f_0 = - \frac{\partial}{\partial p_1}
\end{equation}
The names of these operators $(e_n,\psi_n, f_n)$ borrowed from the theory of simple Lie algebras where we have a triple of (rising operator, Cartan generator, lowering operator) for any node in the Dynkin diagram. Namely, $e_n,f_n$ operators are rising and lowering operators since they add or remove boxes of the Young diagram. Operators $\psi_n$ are similar to Cartan generators and act diagonally in representations.

The same story about cut-and-join operators and Young diagrams translates to other Yangians. We provide here an example of affine super-Yangian $Y(\hat{\mathfrak{gl}}_{1,1})$ that possesses a semi-Fock representation, where vectors are enumerated by super-Young diagrams \cite{Galakhov:2023mak}:
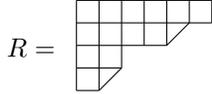
\begin{equation}
	R=\begin{array}{c}
		\begin{tikzpicture}[scale=0.3]
			\foreach \i/\j in {0/-4, 0/-3, 0/-2, 0/-1, 0/0, 1/-3, 1/-2, 1/-1, 1/0, 2/-2, 2/-1, 2/0, 3/-2, 3/-1, 3/0, 4/-1, 4/0, 5/-1, 5/0}
			{
				\draw (\i,\j) -- (\i+1,\j);
			}
			\foreach \i/\j in {0/-3, 0/-2, 0/-1, 0/0, 1/-3, 1/-2, 1/-1, 1/0, 2/-2, 2/-1, 2/0, 3/-1, 3/0, 4/-1, 4/0, 5/0, 6/0}
			{
				\draw (\i,\j) -- (\i,\j-1);
			}
			\foreach \i/\j in {2/-3, 5/-1}
			{
				\draw (\i,\j) -- (\i-1,\j-1);
			}
		\end{tikzpicture}
	\end{array} \hspace{15mm} R = [R_1, R_2, \ldots, R_{l(R)}], \ \text{where} \ R_i \in \frac{1}{2}\mathbb{Z} \ \text{and} \ \ R_1 \geqslant R_2 \geqslant \ldots \geqslant R_{l(R)} > 0
\end{equation}
In case of $R_i$ and $R_{i+1}$ are both half-integers then the condition $R_i > R_{i+1}$ is satisfied.
The numbers $p_s(n,m)$ of super-Young diagrams with $n$ boxes and $m$ half-boxes are collected in the following generating function:
\begin{equation}
	\prod_{k=1} \frac{1 + \eta \cdot q^{k-1}}{1 - q^{k}} = \sum_{n=0} p_{s}(n,m) \, q^n \eta^m =  1 + \eta + q + 2 \eta  q + ( 2 q^2+\eta ^2 q) + 4 \eta  q^2 + \ldots
\end{equation}
As one can see from the generating function, in the case of super-Yangian we need two sets of variables: bosonic $p$-variables $p_a$ (denominator) and fermionic/Grassmann variables $\theta_a$ (numerator). If one assigns degrees $\deg(p_a) = a$ and $\deg(\theta_a) = a - \frac{1}{2}$ then the numbers of homogeneous polynomials of fixed degree are described by the above generating function. 

In the case of super-Yangian the corresponding quiver has two nodes ($(+)$-node and $(-)$-node) therefore the minimal set of operators is twice bigger than in the previous case of $Y(\hat{\mathfrak{gl}}_1)$. Rising/lowering operators:
\begin{align}
	\begin{aligned}
		e_0^{+} &= \theta_1, &\hspace{10mm} f_0^{+} &= \frac{\partial}{\partial \theta_1}, \\
		e_0^-&=\sum\lm_kp_k\frac{\p}{\p\theta_k},
		&\hspace{10mm} f_0^-&=\epsilon_1\epsilon_2\sum\lm_k k\,\theta_k\frac{\p}{\p p_k} \, .
	\end{aligned}
\end{align}
and two super-cut-and-join operators $\hat W^{+}$ and $\hat W^{-}$ \cite{Galakhov:2023mak}:
\begin{equation}
	\begin{split}
		&\hat W^{\pm}=\frac{1}{2}\sum\lm_{a,b= 1}^{\infty}\left[ab\,p_{a+b}\frac{\p^2}{\p p_a\p p_b}-\epsilon_1\epsilon_2(a+b)p_ap_b\frac{\p}{\p p_{a+b}}\right]+\\
		&\quad+\sum\lm_{a,b=1}^{\infty}\left(b - \frac{1}{2} \pm \frac{1}{2} \right)\cdot\left[a\,\theta_{a+b}\frac{\p^2}{\p p_a\p\theta_b}-\epsilon_1\epsilon_2\,p_a\theta_b\frac{\p}{\p\theta_{a+b}}\right]+\\
		&\quad + \frac{\epsilon_1+\epsilon_2}{2}\cdot \sum\lm_{a = 1}^{\infty} \left[a \left(a-\frac{1}{2} \mp \frac{1}{2}\right)p_a\frac{\p}{\p p_a}+ (a-1)\left(a-\frac{1}{2} \pm \frac{1}{2}\right)\theta_a\frac{\p}{\p\theta_a}\right]
	\end{split}
\end{equation}
These super-cut-and-join operators define a new set of polynomials $\mathcal{S}_{R}$, that we call super-Schur polynomials, as the set of own eigenfunctions:
\begin{equation}
	\hat W^{\pm} \mathcal{S}_{R} = w^{\pm}_{R} \mathcal{S}_{R}
\end{equation}
These polynomials have two free parameters $\epsilon_{1,2}$ as in case of usual Jack polynomials, where $\beta$ is a free parameter. Remarkably, super-Schur polynomials also form a orthogonal basis and obey the modified Cauchy identity:
\begin{equation}
	\sum_{R} \frac{\mathcal{S}_{R}\{p, \theta\} \cdot \mathcal{S}_{R}\{\bar{p}, \bar{\theta}\}}{||\mathcal{S}_{R}||^2} = \exp \left\{ \sum_{k = 1}^{\infty} \frac{  p_k \, \bar{p}_k}{k}  + \theta_k \, \bar{\theta}_k \right\}
\end{equation}

\section{Yangian from quiver representations}
\label{Yangain from quiver reps}
Yangians arise as the special algebras that act on the space of BPS states in type IIA string theory with a system of D-branes compactified on the toric Calabi-Yau threefold \cite{Harvey:1996gc}. BPS moduli space can be described in terms of the { \it quiver and superpotential} -- they are extracted directly from the toric Calabi-Yau \cite{Li:2020rij, Ooguri:2009ijd}. What is important for our presentation is that: Yangian acts between different fixed points (of torus action) in the space of quiver representations (= BPS moduli space). The transition amplitudes are given by the equivariant integrals.

\subsection{Definitions: quiver}
\label{def quiver}
Generic quiver data $\fQ$:
\begin{enumerate}
	\item Quiver -- an oriented graph: a collection of nodes $\fQ_0$ + a collection of arrows $\fQ_1$.
	\item Superpotential $W$. We could denote it as $\fQ_2$ since it is made of closed loops in quiver.
	\item Equivariant parameters $\epsilon_{\alpha\in \fQ_1}\in \IC$ assigned to each arrow in $\fQ$ and constrained by a condition that sums of $\epsilon_{\alpha}$'s over any arrow loop contributing to $W$ is zero.
	So that the equivariant weight of $W$ is zero.
\end{enumerate}

Also we adopt the following notations:
\begin{itemize}
	\item $\{a\to b\}$ -- a set of arrows in $\fQ$ flowing from node $a$ to node $b$
	\item $|a\to b|$ -- a number of arrows in $\fQ$ flowing from node $a$ to node $b$
\end{itemize}

\subsection{Quiver representation and torus fixed points}
\label{Quiver representation}
\begin{itemize}
	\item{}
	Quiver representation assigns a vector space to each quiver node and a matrix $B_i$ to each quiver arrow. The dimensions of different vector spaces in quiver nodes could be different, therefore in general $B_i$ are rectangular matrices. Important point is that the matrices $B_i$ are considered up to the change of basis in the quiver nodes.

	\item{}
	Superpotential is defined from the periodic lattice corresponding to torus CY${}_3$ by the following rule:
	\be
	W = \Tr \sum_{\rm faces} (-)^{\rm orientation} \prod_{\rm arrows} B_i
	\ee
	Then we impose so-called F-term equations on matrices $B_i$:
	\begin{equation}
		\frac{\p W}{\p B_i} = 0
	\end{equation}
	These equations correspond to part of famous ADHM equations in the case of CY${}_3 = \mathbb{C}^3$. The other part of ADHM equations (so-called D-term equations) involve additional Fayet-Illiopoulos parameters $\zeta_a$, on which the moduli spaces of solutions can depend in a non-trivial way. In this paper we consider only the case $\zeta_a > 0$ and these D-term equations are irrelevant for our presentation of the simplest Yangian representations. However, for the other choices of parameters $\zeta_a$ the corresponding Yangians can be different, but related by mutations.
	
	\item{}
	Framing is an additional data, which can be represented by additional nodes
	added to the quiver, and it defines a representation of Yangian.
	In particular it can distinguish between $2d$ Fock and $3d$ MacMahon representations.
	
	\item{}
	Torus action scales the matrices $B_i$:
	\begin{equation}
		B_i \to e^{\epsilon_i} \cdot B_i
	\end{equation}
	Vectors of the Yangian representation correspond to fixed points of torus action in the space of quiver representations. In other words, vectors of Yangian representation correspond to a set of matrices $B_i$ (up to change of basis in quiver nodes) obeying F-term equations + fixed point constraints. 
	
	These fixed points of the torus action are labeled by crystals = Young-type diagrams. We further denote these diagrams $\lambda$ and the vector space of the Yangian representation will be space of $\ket{\lambda}$. These diagrams will be constructed from building blocks - atoms $\Box_a$ of different types, that are enumerated by the quiver nodes $a \in \fQ_0$. 
\end{itemize}

\subsection{Yangian algebra}
To construct the Yangian algebra we go through the following algorithm:
\begin{itemize}
	\item Firstly, for each quiver node $a$ we introduce two fields $e^{(a)}(z)$ and $f^{(a)}(z)$. These fields will have a simple interpretation in terms of the diagrams -- they add and remove atoms $\Box_a$ by the following rule:
	\begin{equation}\label{e_weight}
		e^{(a)}(z) \ket{\lambda} = \sum_{\Box_a \in \lambda^{+} } \frac{E_{\lambda, \lambda+\Box_a}}{z - \omega(\Box_a)} \ket{\lambda + \Box_a}
	\end{equation}
	\begin{equation}
		f^{(a)}(z) \ket{\lambda} = \sum_{\Box_a \in \lambda^{-} } \frac{F_{\lambda, \lambda-\Box_a}}{z - \omega(\Box_a)} \ket{\lambda - \Box_a}
	\end{equation}
	There are several comments on the above formulas. As was mentioned in the previous section on Schur polynomials the notation $\lambda^{\pm}$ means the set of atoms outside/insider the diagram $\lambda$ where we can add/remove the atom $\Box_a$ of type $a$. The function $\omega(\Box_a)$ is a weighted coordinate of the $\Box_a$. The weight of the $x_i$ coordinate is $\epsilon_i$ therefore:
	\begin{equation}
		\omega(\Box_a) = \sum_{i} x_{i}(\Box_a) \cdot \epsilon_i
	\end{equation}
	According to this procedure the field actions $e^{(a)}(z) \ket{\lambda}$ and $f^{(a)}(z) \ket{\lambda}$ can only have simple poles and the residues exactly match the vectors $\ket{\lambda \pm \Box_a}$ with one atom added/removed.
	\item Secondly, we should choose the coefficients $E_{\lambda, \lambda+\Box_a}$ and $F_{\lambda, \lambda-\Box_a}$. This choice is important because it controls the commutation relations of the resulting fields $e^{(a)}(z)$ and $f^{(a)}(z)$. In our presentation \underline{{\it the coefficients are extracted from the quiver representation}} described in the previous subsection \ref{Quiver representation}. The extraction procedure goes in four steps:
	\begin{enumerate}
		\item For a diagrams $\lambda$ and $\lambda + \Box_a$ we find a sets of matrices $B_i$ and $B^{'}_i$ satisfying F-term equations + fixed point constraints.
		\item We find a subspace in the space of small perturbations $\Delta B_i$ and $\Delta B^{'}_i$ around fixed point solutions $B_i$ and $B^{'}_i$ that obey F-term relations up to the change of basis in the quiver nodes. From this operation we derive the normalization rules:
		\begin{equation}
			\bra{\lambda} \ket{\lambda} = \text{Eul}_{\lambda}
		\end{equation}
		where the new object $\text{Eul}_{\lambda}$ is the product of all the weights of all matrix coefficients of $\Delta B_i$. The terminology $\text{Eul}$ comes from the fact that the above prescription implicitly corresponds to equivariant (Duistermaat-Heckmann) integration. In that formalism the answer for the integral is given by the sum of the different fixed points of the torus action and each contribution is the Euler class of the tangent space at the fixed point.
		\item At the final step we compute the last ingredients $\text{Eul}_{\lambda, \lambda + \Box_a}$ to compute coefficients $E_{\lambda, \lambda+\Box_a}$:
		\begin{align}
			E_{\lambda, \lambda+\Box_a} &= \frac{\text{Eul}_{\lambda}}{\text{Eul}_{\lambda, \lambda + \Box_a}}
		\end{align}
		To compute coefficients $\text{Eul}_{\lambda, \lambda + \Box_a}$ we should connect two solutions corresponding to diagrams $\lambda$ and $\lambda + \Box_a$ by a linear (generally irreversible) matrix $\tau$ (a zero dimensional analog of the singular Hecke modification \cite{Kapustin:2006pk} shifting the Chern classes of a bundle):
		\begin{equation}
			\label{tau constraint}
			B^{'}_{i} \cdot \tau = \tau \cdot B_{i}
		\end{equation}
		The map $\tau$ is continued to the level of small perturbations and selects a smaller subspace of $\Delta B_i$, $\Delta B^{'}_i$ that obey \eqref{tau constraint} up to the first order. The products of all weights in this smaller subspace defines $\text{Eul}_{\lambda, \lambda+\Box_a}$.
		
		The coefficients $\text{Eul}_{\lambda, \lambda+\Box_a}$ needed for $F_{\lambda, \lambda-\Box_a}$ are computed in the same way but for another pair of diagrams:
		\begin{equation}
			F_{\lambda, \lambda-\Box_a} = \frac{\text{Eul}_{\lambda}}{\text{Eul}_{\lambda, \lambda - \Box_a}}
		\end{equation}
	\end{enumerate}
	According to the above procedure the representation of the Yangian is constructed via simple calculations with matrices. Some details and examples of calculations can be found in \cite{nakajima1999lectures, Rapcak:2018nsl, Rapcak:2020ueh} and \cite[Appendix C]{Galakhov:2020vyb}.
\end{itemize}

\section{The formal definition of Yangian directly from quiver}
\label{Yangain from quiver}


\subsection{Yangian algebra}
\label{formal definition}

$\boxed{\text{\textbf{Yangian $Y_{\fQ}$ is entirely and directly defined by the quiver $\fQ$.}}}$

For each vertex $a$ in $\fQ$ we associate the generators:
positive and negative "simple roots" $e^{(a)}(z), f^{(a)}(z)$ and related "Cartan elements" $\psi^{(a)}(z)$ :
\begin{align}\label{series}
	e^{(a)}(z) = \sum_{n=0}^{\infty}  \frac{e^{(a)}_n}{z^{n+1}}, \hspace{10mm}
	\psi^{(a)}(z) = \sum_{n=-\infty}^{\infty}  \frac{\psi^{(a)}_n}{z^{n+1}}, \hspace{10mm}
	f^{(a)}(z) = \sum_{n=0}^{\infty}  \frac{f^{(a)}_n}{z^{n+1}}
\end{align}

$Y_{\fQ}$ is a superalgebra.
Therefore the root generators acquire definite $\IZ_2$-parity $P$: bosonic $P=0$ or fermionic $P=1$.
Parity of node $a$ is defined according to the following formula:
\begin{equation}
	P_a=(|a\to a|+1)\; \mod \;2\,.
\end{equation}
Cartan generators are all bosonic.

Super-commutation relations are defined by the maps/arrows of $\fQ$
\begin{equation}
	\begin{split}
		\psi^{(a)}(z) \psi^{(b)}(w) &\cong  \psi^{(b)}(w) \psi^{(a)}(z)\,,\\
		e^{(a)}(z) e^{(b)}(w) &\cong   (-1)^{P_aP_b}\varphi^{\fQ}_{ab}(z-w)\;e^{(b)}(w) e^{(a)}(z)\,,\\
		f^{(a)}(z) f^{(b)}(w) &\cong   (-1)^{P_aP_b}\varphi^{\fQ}_{ab}(z-w)^{-1}\; f^{(b)}(w) f^{(a)}(z)\,,\\
		\psi^{(a)}(z) e^{(b)}(w) &\cong  \varphi^{\fQ}_{ab}(z-w)  e^{(b)}(w)\; \psi^{(a)}(z)\,,\\
		\psi^{(a)}(z) f^{(b)}(w) &\cong   \varphi^{\fQ}_{ab}(z-w)^{-1}  f^{(b)}(w)\; \psi^{(a)}(z)\,,\\
		\left[e^{(a)}(z),f^{(b)}(w)\right\} &\cong-\delta_{ab}\frac{\psi^{(a)}(z)-\psi^{(b)}(w)}{z-w}\,.
	\end{split}
\end{equation}
Where:
\begin{itemize}
	\item $\left[x,y\right\}:=xy-(-1)^{P_xP_y}yx$ is a supercommutator
	\item Sign $\cong$ equates Taylor series expansion at the points $z=\infty$, $w=\infty$ on both sides
	up to the terms of the  form $z^{n\geq 0}w^m$ and $z^{n}w^{m\geq 0}$
	\item Quiver bond factors:
	\begin{equation}
		\varphi^{\fQ}_{ab}(u):=\frac{\prod\lm_{\alpha\in \{a\to b\}} (u+\epsilon_\alpha)}{\prod\lm_{\beta\in \{b\to a\}} (u-\epsilon_\beta)}
	\end{equation} 
\end{itemize}

{\bf Remark 1:} Cubic and higher order Serre relations
require additional consideration, left beyond the scope of the present text.
A conjecture for Serre relations in the case of a generic quiver is given in \cite{Negut:2022pka}.
Cases $Y(\widehat{\fg\fl}_{m|n})$ are described in \cite{bezerra2019braid}.
Some suggestions for $Y(K_{\IP^2})$ and $Y(K_{\IP^1\times \IP^1})$ (not based on Lie algebras) are given in Appendix D of \cite{Galakhov:2021vbo}.

{\bf Remark 2:} Canonically Yangians with different shifts (when one allows for non-zero negative Cartan element modes $\psi^{(a)}_{-\fs\leq n <0}\neq 0$) are considered to be different algebras \cite{Kamnitzer2012YangiansAQ, Kodera:2016faj}.
However, physically, one might try to classify them as simply different representations.
Both treatments have their own peculiarities.
On one hand mere framing modification leaving the unframed quiver diagram intact introduces shifts \cite{Galakhov:2021xum, Noshita:2021dgj}.
And the naive $n^{\rm th}$ tensor power on the Yangians could be also installed in the physical picture as a framing modification, thus mixing together of Yangians with different shifts in a single object of a tensor category is not forbidden.
Moreover, for the quiver Yangians not based on the affine Lie superalgebras the shift $\fs$ is infinity, formally speaking, it can not be bounded above for infinite MacMahon-like crystal representations \cite{Galakhov:2022uyu}.
On the other hand, the negative Cartan element modes seem somewhat decoupled from the algebra since in the Yangian relation $\left[e_n^{(a)},f_m^{(b)}\right\}=\delta_{ab}\psi^{(a)}_{n+m}$, $n,m\geq 0$ only non-negative modes are generated.
This induces certain obstacles in a search for the universal (representation independent) co-product structure that is a homomorphism of algebras.

\section{Yangian and quantum field theory}
\label{Yangain from QFT}
Finally, we could name another seemingly different source of the Yangian algebra as a multi-dimensional supersymmetric QFT.
The crucial difference of this approach with mentioned above is that we do incorporate an effective theory of D-branes in the type IIA string theory compactified on the toric Calabi-Yau.
The effective QFT in this picture is related to the ADHM construction and instanton equations in diverse dimensions only implicitly after a careful analysis of D-brane charges and theory moduli \cite{Douglas:1996sw}.

Nevertheless, this physical approach to a mathematical problem of finding algebraic structures reveals richness and fruitfulness repeating in some aspects the story about the physical avatar of the knot invariant construction problem -- the 3d Chern-Simons theory and the boundary conformal WZW theory \cite{Witten:1988hf}.

Here we mention this story in a seemingly inverse order.
The Yangian algebra admits a Hopf algebra structure, in particular, one could define the co-product even for the affine cases \cite{GUAY2018865, 2019arXiv191106666U, Bao:2023kkh}.
Having the co-product $\Delta$ we can restore the R-matrix -- the intertwining operator permuting the order of multipliers in a tensor square:
\begin{equation}
	\Delta_{12} R = R \Delta_{21}
\end{equation}
Both the co-product and the R-matrix depend on the complex spectral parameter having the same nature as the generating parameter in the series \eqref{series}, or the equivariant weight in \eqref{e_weight}.
$R$-matrix is a solution of the Yang-Baxter equation ensuring  that the representation tensoring structure is indeed associative.

On the other hand the R-matrix, a transfer matrix may be derived independently of the Yangian structure starting with spin chains and other integrable models like the Calogero model \cite{Awata2000}.
The relation between the QFTs and integrable models is well-known in the literature \cite{Nekrasov:2009uh, Nekrasov:2009ui} as the gauge/Bethe correspondence.

The physical approach allows one to argue the appearance of such non-trivial relations as the Yang-Baxter equation in a quite elegant way: two sides of the equality are just different yet homotopic paths in the parameter (moduli) space of the QFT, then the equality of two sides is solely an absence of hysteresis in the theory.
The absence of hysteresis for theories in question is a rather natural property induced by supersymmetry.
On the other hand this approach makes the other avatars of the Yangians mentioned above rather obscure.
This story becomes even more intriguing \cite{Galakhov:2022uyu} when a naive attempt to restore the whole chain of relations underlying the gauge/Bethe correspondence between QFT vacua and solutions to the Bethe equations including the construction of the co-product and the R-matrix fails.

Nevertheless, the R-matrix evolution (actually, in the both cases of the Yangian and quantum algebras and their mixtures like quantum toroidal algebras) in the QFT can be expanded in a series \cite{Dedushenko:2021mds, Bullimore:2021rnr, Dedushenko:2023qjq, Crew:2023tky, Galakhov:2020upa, Galakhov:2014aha} of transitions through instanton/soliton jumps between overlapping levels of effective QFTs.
This observation allows us to expect that the algebra of BPS solitons/instantons reproduces the Yangian in an inexplicit way, in other words we expect to have the following relations:
\begin{equation}
	E_{\lambda,\lambda_+\Box}=\lim\lm_{T\to \infty}\langle\lambda+\Box|e^{\frac{\I}{\hbar}T\,H}|\lambda\rangle\,,
\end{equation}
for some effective Hamiltonian $H$.

\section*{Acknowledgements}

Our work is partly supported by grants RFBR 21-51-46010 ST\_a (D.G., A.M., N.T.), by the grants of the Foundation for the Advancement of Theoretical Physics and Mathematics “BASIS” (A.M., N.T.) and by the grant of Leonhard Euler International Mathematical Institute in Saint Petersburg (D.G., N.T.).

\bibliographystyle{utphys}
\bibliography{biblio}

\end{document}